\documentclass[sigconf]{acmart}
\usepackage{lipsum}
\usepackage{array}
\usepackage{subcaption} 
\usepackage{multirow}
\usepackage{makecell}
\usepackage{fontawesome}
\usepackage{tabularx}
\usepackage{graphicx}
\usepackage{placeins}
\usepackage{booktabs}
\usepackage[export]{adjustbox}
\usepackage{cellspace}
\usepackage{adjustbox}
\usepackage{tabularray}
\usepackage{footnote}
\usepackage[capitalize]{cleveref}
\usepackage{multirow}
\usepackage[multiple]{footmisc}
\usepackage{hyperref}
\usepackage{enumitem}
\usepackage[normalem]{ulem}
\useunder{\uline}{\ul}{}

\AtBeginDocument{%
  }

\setcopyright{acmlicensed}
\copyrightyear{2025}
\acmYear{2025}
\acmDOI{XXXXXXX.XXXXXXX}

\definecolor{vividauburn}{rgb}{0.58, 0.15, 0.14}
\definecolor{rufous}{rgb}{0.66, 0.11, 0.03}
\definecolor{red-brown}{rgb}{0.65, 0.16, 0.16}
\definecolor{red(ncs)}{rgb}{0.77, 0.01, 0.2}

\begin{document}

\title{AnyAni: An Interactive System with Generative AI for Animation Effect Creation and Code Understanding in Web Development}

\author{Tianrun Qiu}
\orcid{0009-0002-9878-663X}
\affiliation{%
  \institution{Southern University of Science and Technology}
  \city{Shenzhen}
  \state{Guangdong}
  \country{China}
}
\email{qiutr@mail.sustech.edu.cn}

\author{Yuxin Ma}
\authornote{Corresponding author.}
\orcid{0000-0003-0484-668X}
\affiliation{%
  \institution{Southern University of Science and Technology}
  \city{Shenzhen}
  \state{Guangdong}
  \country{China}
}
\email{mayx@sustech.edu.cn}


\begin{abstract}
Generative AI assistants have been widely used in front-end programming. However, besides code writing, developers often encounter the need to generate animation effects. As novices in creative design without the assistance of professional designers, developers typically face difficulties in describing, designing, and implementing desired animations. To address this issue, we conducted a formative study (N=6) to identify the challenges that code developers face when dealing with animation design issues. Then, we introduce AnyAni, a human-AI collaborative system that supports front-end developers in the ideation, manipulation, and implementation of animation effects. The system combines the assistance of generative AI in creative design by adopting a nonlinear workflow for iterative animation development. In addition, developers can understand and learn the code generated for implementing animations through various interactive methods. A user study (N=9) demonstrated the usability of AnyAni in animation effect creation support for developers.
\end{abstract}


\begin{teaserfigure}
 \includegraphics[width=\textwidth]{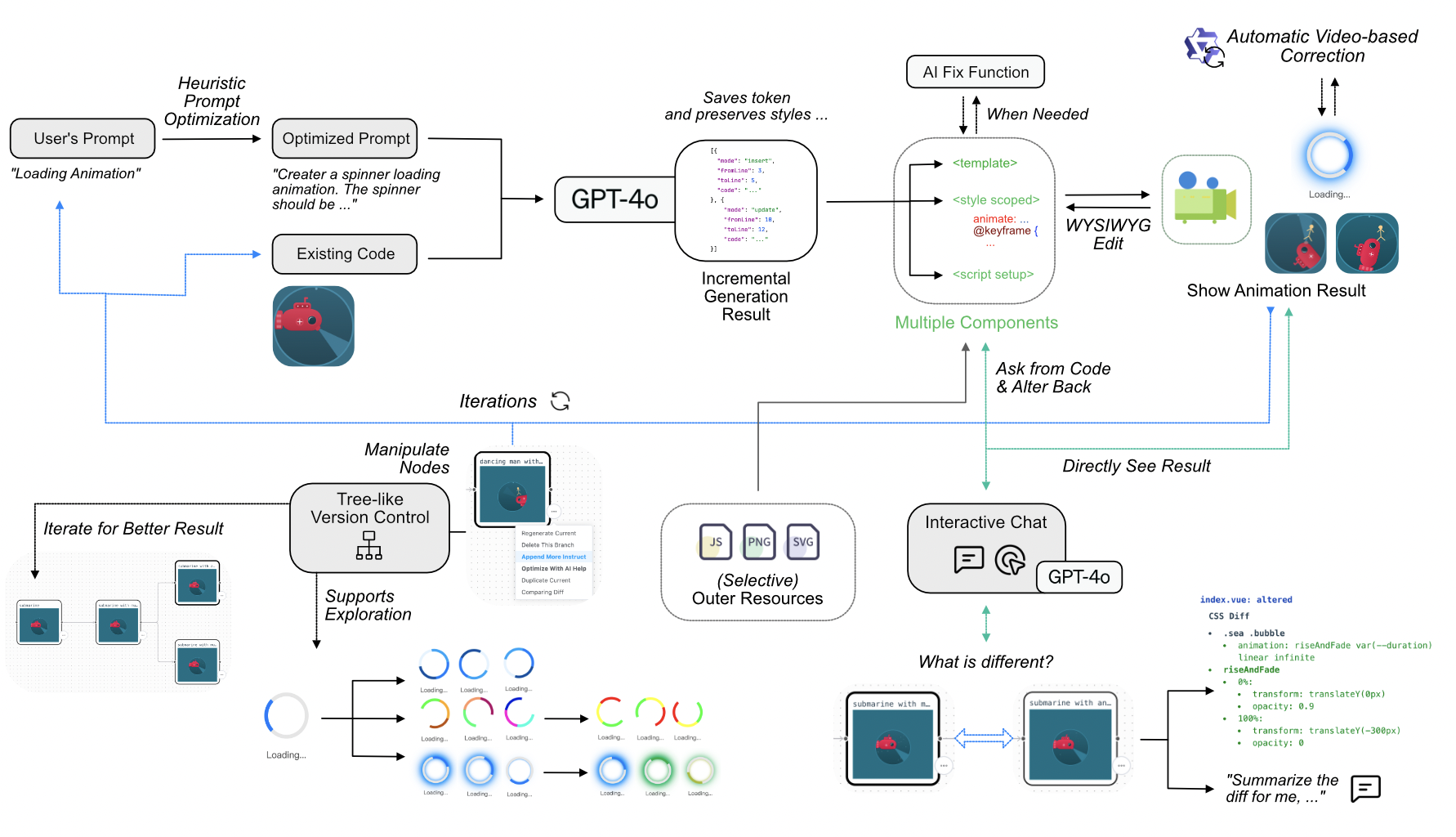}
 \caption{We introduce AnyAni, a human-AI collaborative system that leverage large language models to empower users to create, explore, iterate and understand web animation effects.}
 \Description{This image illustrates the workflow and output of AnyAni, a human-AI collaborative system designed to leverage large language models to empower users in creating, exploring, and understanding web animation effects. Users input a prompt and receive heuristic prompt optimization, with the option to input existing relevant code. The prompt and code are then submitted to the GPT-4o model, which generates incremental results. These results are parsed into multiple components and ultimately displayed in the Show Panel, where users can engage in WYSIWYG editing. Additionally, users can perform iterative and exploratory programming via tree-structured version management, automatic video-based correction and interpret code through the Interactive Chat Panel. This Chat Panel allows users to decode the code and compare differences between generated nodes, facilitating a more comprehensive and interactive development experience.}
 \label{fig:teaser}
\end{teaserfigure}


\maketitle

\section{Introduction}

In the realm of web development, the role of animation and visual elements in crafting an engaging user experience is significant. Animations serve as an important way for enhancing interactivity, guiding user attention, and providing visual feedback, thus playing a pivotal role in the overall user interface and user experience design; thus designing animation is also an important realm of computer-aided design (CAD)~\cite{10.1145/108844.108906, zhang1999will, chang1993animation, Hao2021DigitalMO, Storey2012}.

In development paradigm particularly within domains such as full-stack development and the work of independent developers, there exists a multifaceted role that encompasses not only the technical aspects of development but also the aesthetic design of web elements, ranging from creating complicated page animations refining subtle visual effects. However, the process of conceptualizing, drafting, and implementing these animations presents a significant challenge for developers, especially those lacking a foundation in creative design. The challenges encountered in the design and application of animations can be attributed to several factors. First, the absence of professional training in visual design among developers may result in inadequately articulating the requirements of animation design. As such, developers may prefer to use broad, imprecise terminology when describing animations~\cite{10.1145/3613904.3642812}, which poses difficulties in translating these descriptions into tangible visual elements and corresponding source code. Furthermore, the inherent complexity associated with animation features in web development, such as Cascading Style Sheets (CSS) and animation APIs in JavaScript, aggravates the learning curve. This complexity can act as a substantial obstacle to efficient development practices~\cite{Storey2012}. Developers may find themselves buried by extensive searching of documentations to grasp the functionalities of specific parameters during the implementation phase.

This study seeks to explore the design space and workflow associated with the development of web animation effects. Subsequently, it aims to apply these design principles to develop an AI-assisted tool that enhances the creation, iteration, and comprehension of web animation effects. By interviewing six experienced front-end developers, we identified their main workflows in creating web animations, their views on the capabilities and usages of current large language models (LLMs) in this field, and the major difficulties and pain points they encounter in their development and learning processes. Building on the key findings of our formative study, we developed AnyAni, an AI-integrated system that combines large language models with various interaction methods, aiming to assist users in creating web animation effects more efficiently and effectively, while also helping them understand code and expand their ideas.

The AnyAni system can generate high-quality web animation effects based on user-provided prompts and existing code snippets through multi-component and incremental generation methods. It also incorporates collaborative iterations between the large model and human developers. Additionally, when users need cognitive support, the system can help them analyze and understand code through the interactive chat panel. To evaluate the effectiveness of the AnyAni system in practical applications, we conducted a user study with nine participants, designed two tasks to compare the system with a baseline system, and conducted semi-structured interviews. User feedback revealed the effectiveness of AnyAni in generating and optimizing web animation effects while enhancing code comprehension. The results of the NASA-TLX analysis \cite{nasa-tlx} further proved that AnyAni outperformed the baseline system in multiple dimensions and significantly enhanced users' performance.

In summary, our work offers the following contributions:

\begin{enumerate}
    \item \textbf{Formative Study:} By interviewing six front-end development developers with different levels of animation development, we confirmed common workflows in web animation creation and identified prevalent pain points when performing these tasks and utilizing large models for assistance.
    \item \textbf{System Prototype:} We developed a comprehensive system, AnyAni, which integrates the entire workflow, enabling developers to conveniently conceptualize, explore, generate, iterate, and understand web animation effects.
    \item \textbf{Generation Techniques and Prompts:} We explore to employ incremental and multi-component generation techniques to better achieve generation, exploration, and iteration in creative coding.
    \item \textbf{User Study:} Through dual-task evaluation and baseline comparison, we demonstrated the usability of the AnyAni system in various contexts. We also provided an in-depth analysis of how users interacted with our system.
\end{enumerate}
\section{Related Works} 
\subsection{Web-based Animation Effects Design} \label{anim-tools}
Animation effects are dynamic transformations of visual elements over time, which aim at improving interactivity and user experience across web and multimedia platforms~\cite{chang1993animation}. In contrast to static elements, animation effects enhance attention and comprehension by leveraging dynamic transitions and interactive features to convey information more efficiently and effectively~\cite{paas2007instructional,wu2020predicting}. The ability to enrich visual engagement and interactivity has driven the growing utilization of animation effects in various fields, including augmented reality~\cite{xia2023realitycanvas}, data visualization~\cite{cao2023dataparticles}, and data storytelling~\cite{lan2021kineticharts}. By progressing through two phases, concept initialization and implementation exploration, designers can ensure consistency between animation effect design and functionality, resulting in more engaging and user-centered outcomes.

The process begins with concept initialization, where design objectives, requirements, and constraints are established through collaborative discussions, ensuring a comprehensive understanding of user needs and forming the foundation for the design framework. In this phase, Large Language Models (LLMs)~\cite{BrownMRSKDNSSAA20,gpt4} are frequently employed to offer initial recommendations and refine requirements through conversational interactions. However, without proper tools, its assistance can be limited~\cite{10.1145/3563657.3596014}.

The exploration phase follows, characterized by the implementation, continuous iteration and refinement of design solutions. While designers often use professional computer graphics tools (CST) such as Adobe After Effects~\footnote{\url{https://www.adobe.com/products/aftereffects.html}} and Apple Motion~\footnote{\url{https://www.apple.com/final-cut-pro/motion}}, they are predominantly complex commercial software with a steep learning curve, and have limited ability to create interactive animations. Front-end Developers, on the other hand, use CSS animation or web animation libraries, such as Anime.js~\footnote{\url{https://animejs.com}}, Motion~\footnote{\url{https://motion.dev}}, GSAP~\footnote{\url{https://gsap.com}} and Velocity.js~\footnote{\url{http://velocityjs.org}}. However, these techniques are relatively not intuitive, and their APIs can be complicated for mastering.

Recent research has begun to address this issue by developing end-to-end tools for authoring web animation effects. For instance, Cao et al.~\cite{cao2023dataparticles} introduced a natural language-based system to explore unit visualization animations and rapidly prototype data storytelling. Similarly, Liu et al.~\cite{liu2024logomotion} employed LLMs to assist non-professional designers in creating animated logos and generating corresponding animation code. Moreover, the DynaVis approach by Vaithilingam et al.~\cite{vaithilingam2024dynavis} enabled fine-grained style editing and customization through dynamically generated widgets. These studies aim to bridge the gap between design and implementation, offering new solutions for the authoring of web animation effects in a more integrated and streamlined manner. Creative version control systems, such as work by Nicholas et al.~\cite{nicholas2022creative}, Micro-Versioning~\cite{mikami2017micro} and Spellburst~\cite{10.1145/3586183.3606719}, also play a critical role in documenting and tracking each iteration, facilitating comparative analysis and informed decision-making. Researchers also discussed methods to facilitate AI to help learning creative programming~\cite{Jonsson2022Jun}.

However, existing animation effect creation tools~\cite{liu2024logomotion,cao2023dataparticles,10.1145/3586183.3606719} often focus on limited aspects of the iteration process. For example, LogoMotion only supports limited animation forms for the specific scene of logo design, can only convert predefined static elements into dynamic effects and only support linear, language-based modification. In the meantime, DataParticles relies on preset templates, thus limiting its usage scope. Spellburst is a trailblazing work discussing about creative version control paradigm, but it only support creating simple visual effects, and its lack of code consistency support and code understanding assistance, the handling of malfunctioning code and guidance for further exploration resulted in a higher cognitive load~\cite{10.1145/3586183.3606719}.

Instead, our framework offers an approach that spans the entire workflow and breaks these limitations by offering users the freedom to choose any animation elements and effects they wish to implement. It leverages LLMs to generate recommendations from vague prompts to aid in brainstorming and decision-making, and is capable to build the whole animation from scratch. A node-based version control panel enables users to visualize and track design exploration, while real-time feedback of manual operations on both code and extracted animation properties allows for precise and flexible adjustments of style parameters. The incremental generation feature ensures consistency while preventing model laziness, combining with code understanding assistance, makes the system well-suited for handling complex animations. 

\subsection{Generative AI for Design and Programming}
Generative AI, with its powerful capabilities and widespread adoption, has significantly lowered the barriers in traditionally high-threshold fields such as design and programming, which once required substantial learning efforts and specialized knowledge. In the field of design, collaborations with generative AI have been extensively applied in image creation~\cite{ramesh2021zero,borji2022generated,qiao2019mirrorgan}, graphic design~\cite{feng2023promptmagician,chiou2023designing}, and visualization~\cite{cao2023dataparticles,wang2023llm4vis}. It can also generate animation videos~\cite{videoworldsimulators2024, chen2025goku, lin2025omnihuman1}, but none of these generated animations can be edited and are not web-based. In programming, generative AI has contributed not only to code completion~\cite{gpt4,roziere2023code}, which primarily benefits professional developers by improving development efficiency, but also to code understanding assistance~\cite{nam2024using,kazemitabaar2024codeaid} and code generation~\cite{cheng2024biscuitscaffoldingllmgeneratedcode}, enabling non-expert users to accomplish end-to-end tasks without writing code or learn programming faster~\cite{10.1145/3544548.3580919}.

While generative AI has demonstrated remarkable capabilities in content creation, achieving optimal results often requires iterative prompt refinement and multi-round interactions to optimize the model's output to meet user expectations~\cite{liu2022design, 10.1145/3597503.3608128}, and prompting can be tricky for novice programmers~\cite{10.1145/3617367}. Additionally, the controllability of generative AI remains limited, making it challenging to perform fine-grained customization. Our framework addresses these limitations by not only leveraging the strengths of generative AI in design and programming to accomplish the complex task of creating end-to-end web animation effects but also by providing comprehensive support throughout the workflow. 
\section{Formative Studies}

In this section, we present a formative study conducted to assess the level of difficulty and the challenges that common programmers encounter when generating dynamic effects for web pages. Additionally, we evaluate their existing use of LLMs for code generation for their professional and learning scenarios. Our objective is to deduce insights on how their current workflows can be optimized through the findings of this study.

\subsection{Participants and Method}

We conducted semi-structured interviews with six developers who possessed extensive experience in the domain of front-end development. Each of the respondents had a minimum of six years' experience in front-end development. Their roles comprised three researchers (N=3, U1, U2, U3), two graduate students (N=2, U4, U6), and one software engineer (N=1, U5). Five out of the six participants (representing 83\%) have frequently applied generative AI assistants in their daily programming activities, while one participant (accounting for 17\%) has used them only occasionally, but have often observed how their colleagues use LLMs to facilitate development. All of them reported that they have had experience in developing web animations and interaction effects to fulfill certain requirements. Three of them are highly experienced (U1, U2, U6) in web animation development while other three interviewees are novices in this subfield. Although our framework is primarily geared towards people with less experience, interviews with experts may still be important because they can share the best practices they suggest and recall the key points of difficulty when learning web animation in the first place, which may inspire our research. Each semi-structured interview lasted between 30 to 45 minutes. Participants were recruited through word-of-mouth and were not compensated for their involvement.

The formative interviews began with questions about the participants' roles and their backgrounds in front-end development. Then, we surveyed about their workflows, favorite development tools and their challenges faced when learning about and developing with front-end technology stacks and web animation effect libraries. Following this, we delved into their experiences concerning the utilization of LLMs to empower the aforementioned contexts, as well as the inconveniences encountered in the process and their unmet needs.

\subsection{Findings and Design Goals}

\subsubsection{Requirements of animation design and production} \label{form-1}

Many participants (50\%) in our study indicated that there is a high learning curve for mastering the creation of web animation effects and complex styles. Participant U2 pointed out that \textit{``If you're not specifically learning CSS or other libraries, in fact, there are many animation effects that you cannot implement.''} Similarly, Participant U3 and U5 reported that their lack of proficiency in this domain resulted in a higher time cost during their programming process when they were learning. 

Participants also observed that this issue partially arises from the complexity of the associated APIs and parameters. For instance, U2 and U4 remarked that \textit{``the animation APIs lack an intuitive interaction mechanism.''} Furthermore, U1 highlighted the challenge of achieving the desired outcome, noting that \textit{``There are times when I have written the code, but the result does not match my expectations.''} 

Consequently, U5 also noted that this type of work is often assigned to a dedicate designer instead of the programmer because of its complexity, Furthermore, U1 and U6 suggested that there is a desire to leverage the capabilities of AI to integrate these two roles into a more efficient workflow that can be completed by one programmer.

\subsubsection{Learning, understanding and analysis} \label{form-2}

Some participants (N=2) consider existing LLM-based services may not provide satisfactory answers and solutions to some complex designing issues without proper guidance. U4 expressed, \textit{``When it comes to more profound questions, you cannot expect that they can provide satisfying solutions instantly.''} Issues related to confusion or absence of context also impact the resolution of user inquiries. \textit{``AI are sometimes confused about what you are asking about now, so it may provide irrelevant or disordered answers,''} said U1.

Moreover, after specific code was generated by LLMs, many participants reported that understanding the entire code structure and details can be time-consuming, citing that it may also be challenging for personal learning when engaging with these systems. U2 complained that the verbosity of output messages hampers his ability to concentrate on the desired explaining content, so he tend to focus directly on the code generated, potentially slows down his understanding. U4 indicated that due to the unfamiliar and ambiguity of parameter values, merely observing the result code might not be sufficient for thorough comprehension. 

Participants also reported that after they implemented the code suggestions made by AI, they encountered difficulties in assessing which parts of the content were modified, resulting in an inefficient analysis process.

\subsubsection{AI-generated results not yet production-ready} \label{form-3}

Participants also complained that current LLMs are not directly capable for creating ready-to-use web animation effects and other kinds of complex style sheet effects. As U4 described: \textit{``sometimes result in the generation of overly simplistic and generic demos when attempting to create certain effects.''} Additionally, owing the the complexity of animation APIs, \textit{``you cannot expect AI to produce perfect results''} (U2), which also reflects the irreplaceability of human beings in this task.

Moreover, creative coding usually requires multiple iterations to achieve improved results by its nature \cite{10.1145/3586183.3606719, 10.1145/3613904.3642812}, but current LLMs often lack the ability to fulfill this demand. U1 pointed out that \textit{``when you modify your requirements, LLMs often confuse them with original ones,''} which may result in failure to meet the new requirements or preserve the existing modifications. Besides, U6 noted that in his design practice, many code LLM services frequently exhibit laziness, which means they consequently omit many details after several rounds of conversation; this type of laziness can also lead to changes in design style, as noted by U1 and U6.

Due to these reasons, many participants expressed a desire for a tool that could realize the vision where \textit{``animation effect evolves according to my textual narration with ease, resulting in AI-generated content that is gradually aligned with my concept''} (word of U1).

\subsubsection{Difficulty in articulating requirements} \label{form-4}

Many participants also reported feeling frustrated when trying to articulate their precise needs to LLMs. U3 pointed out that \textit{``Often I find it hard to present [my needs] clearly, and thus am unable to search or ask about them.''} The lack of relevant proper names for certain animation effects is also a major reason for this frustration (U5). 

In addition, U4 noted that when writing prompts, there are times when their personal ability is insufficient to fully articulate the needed prompt. Therefore, they have to open another large model dialogue window or use a search engine to gain inspiration. Otherwise, the prompts written often fail to accurately guide the model to generate responses and results that align with the user's intentions, requiring an additional round of conversation to provide further instruction so the model can truly understand the user's thoughts.

\section{Design Considerations}

After the formative study, we have identified the following two goals that our design of the system should achieve:

\begin{enumerate}
    \item \textbf{G1.} To help programmers implement their concepts related to animation effects in more interactive, intuitive and accurate ways.
    \item \textbf{G2.} To better assist programmers in understanding the code generated by LLMs and the principles behind achieving specific animation effects.
\end{enumerate}

To meet these objectives, our system adheres to four key design principles derived from formative study insights:

\begin{enumerate}
    \item \textbf{DP1.} \label{dp-1}\textit{Capable of helping programmers realize ideas about animations and dynamic effects}: Supports non-expert users in creating web animations via human-in-the-loop co-creation with multi-role AI agents, mitigating steep learning curves.
    
    \item \textbf{DP2.} \label{dp-2} \textit{Provide adequate support for code comprehension and principles learning}: Addresses LLM-generated code opacity and linear workflow limitations through contextual explanations, interactive learning aids~\cite{cheng2024biscuitscaffoldingllmgeneratedcode}, and nonlinear exploration.  

    \item \textbf{DP3.} \label{dp-3} \textit{Support human-computer collaborative and iterative optimization of code}: Enhances LLM-based code generation via structured human-computer collaboration~\cite{Kantosalo2016ModesFC}, enabling joint refinement of outputs through iterative feedback loops.  

    \item \textbf{DP4.} \label{dp-4} \textit{Help user express ideas in an easier way}:  Translates abstract animation concepts into concrete specifications via visual scaffolding, prompt templates, and domain-specific syntax and API simplification.
\end{enumerate}
\section{System}

AnyAni is designed as a web-based system that utilizes LLMs and interactive visual interface to assist users in quickly implementing and exploring web animation effects, as well as helping users to understand and learn. As mentioned in DP\ref{dp-3}, the system enables users to directly intervene and modify the code, facilitates incremental generation, offers generation based on multiple components, and supports tree-based non-linear versioning. Furthermore, in terms of code generation, we devoted efforts to enhance the quality and alignment with requirements of the generated results, and strove to mitigate issues such as model laziness and inconsistent styles. 

In the following, we will analyze the system layout, core features and implementation technology stack of AnyAni system.

\subsection{System Layout}

\begin{figure}[htb]
    \centering
    \includegraphics[width=1\linewidth]{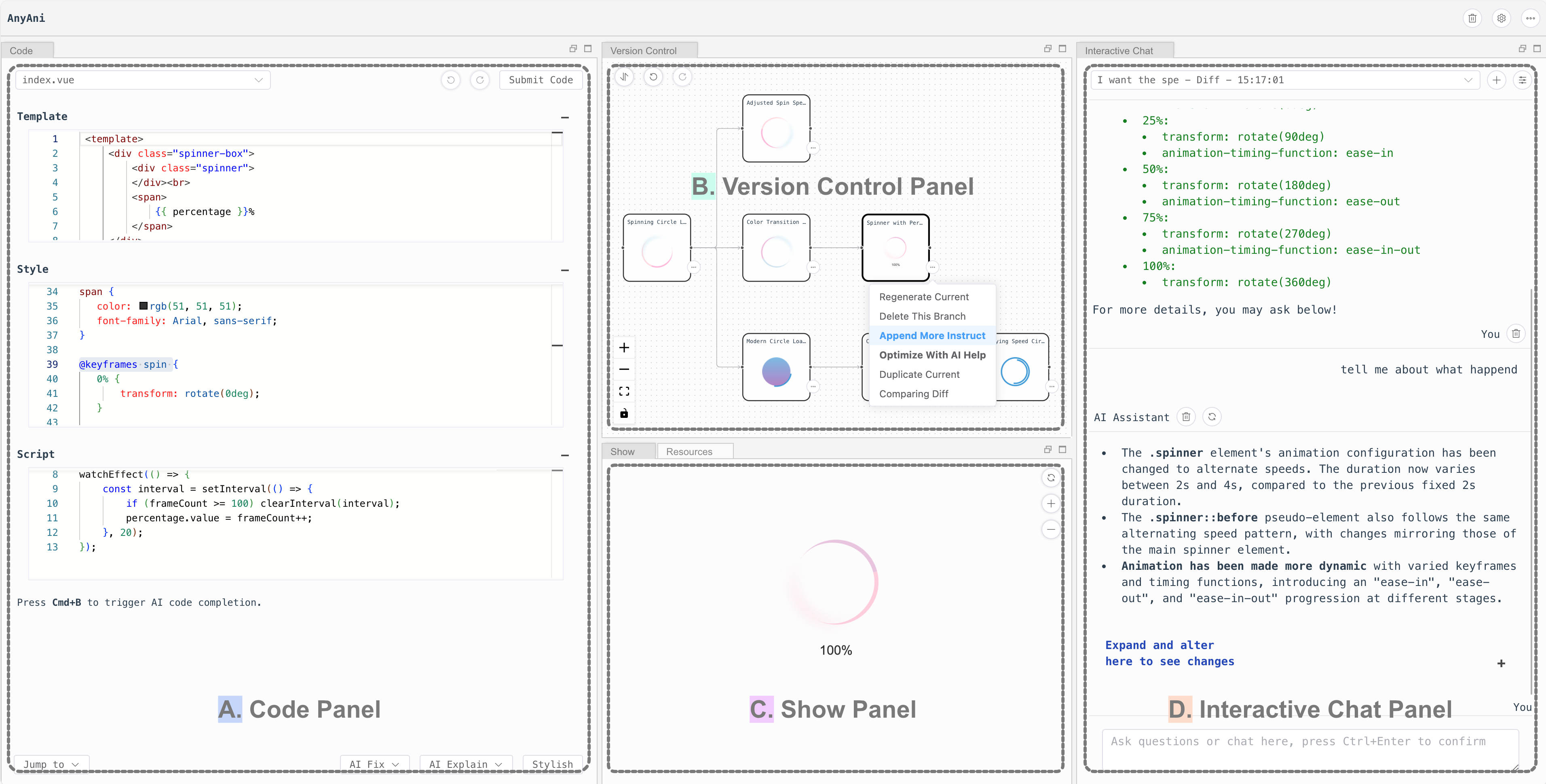}
    \caption{Overview of AnyAni's user interface, which consists of four organically integrated and freely adjustable main panels: (A) Code Panel, (B) Version Control Panel, (C) Show Panel and the (D) Interactive Chat Panel}
    \label{fig:layout}
\end{figure}

\begin{figure}[htb]
    \centering
    \includegraphics[width=1\linewidth]{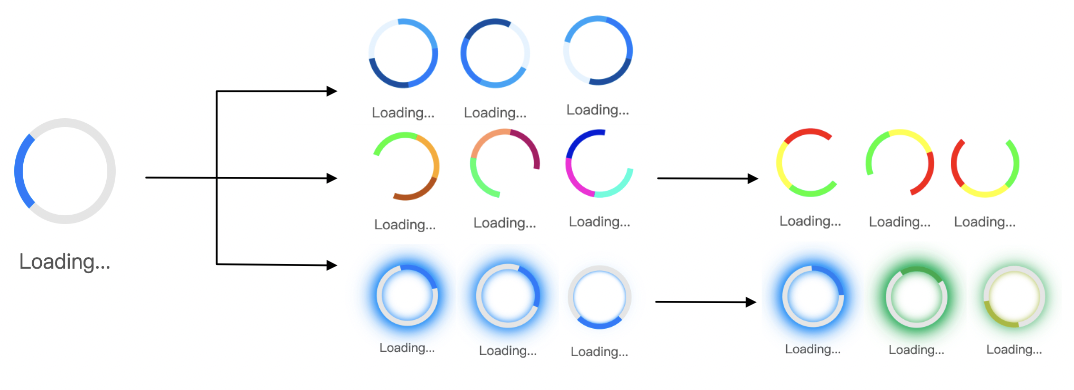}
    \caption{Example of non-linear exploration on the task of loading animation}
    \label{fig:example-1}
\end{figure}

\begin{figure}[htb]
    \centering
    \includegraphics[width=0.72\linewidth]{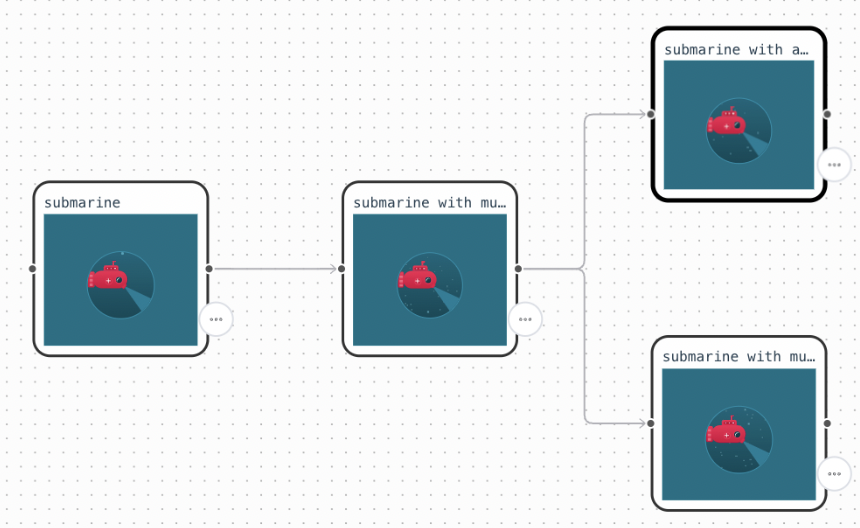}
    \caption{Example of iterative optimization on a submarine animation, from a stable submarine, to submarine with multiple stable bubbles, and to submarine with dynamic bubbles, or submarine with a dynamic front light}
    \label{fig:example-2}
\end{figure}

As shown in Fig.~\ref{fig:layout}, the main layout of AnyAni consists of four integrated parts: Code Panel, Version Control Panel, Show Panel and the Interactive Chat Panel. All panels interacts and synchronizes with each other, and can be freely rearranged, resized and reconfigured. Examples of generated results are shown in Fig.~\ref{fig:example-1} and Fig.~\ref{fig:example-2}.

\begin{itemize}
    \item \textbf{The Code Panel} consists of three separate code boxes, i.e., template, style, script, which corresponds to the three main parts for implementing HTML-based web pages.
    It is worth-noting that many front-end frameworks, such as Vue~\footnote{\url{https://vuejs.org/}} and React~\footnote{\url{https://react.dev/}}, also include such code organization into the three parts and can be supported in our system.
    Users can select the current web page filename in the top left corner. When the user clicks ``submit code,'' all user changes will be instantly synchronized to the Show Panel and can be inherited by all subsequent nodes. When editing, one can also utilize the power of AI code completion.
    \item \textbf{The Version Control Panel} employs a hierarchical non-linear architecture to facilitate dynamic dialogue management in creative coding tasks, where each node represents a discrete input-output interaction round (either AI-generated or manually modified). Built upon Vue-Flow's framework, the system enables intuitive node manipulation through drag-and-drop functionality, with each node embedded in an iframe for visual comparison. This structure automatically constructs conversational histories through parent-child node relationships while supporting branch-based development, allowing new prompts to merge selectively with specific historical contexts. Key functionalities include context-aware operations (prompt appending, AI-assisted optimization, duplication, diffing, and deletion) performed at any node, with branch isolation ensuring contextual coherence. The panel synchronizes bidirectionally with the Show Panel for real-time visualization, implementing a version-controlled chat history system that simultaneously addresses nonlinear versioning requirements and long-context preservation challenges in generative workflows.
    \item \textbf{The Show Panel} contains the visual result of current node, and synchronizes with all other panels.
    \item \textbf{The Interactive Chat Panel} allows users to ask questions freely with context, including asking about specific lines of code. Users are also able to view LLM's explanations and optimizations of the code, check the difference information between another node, and further inquire about the results of the above tasks. Meanwhile, the system automatically parses all relevant parameters and enables users to interactively modify these parameters in the same panel, facilitating a ``learning by doing''~\cite{cheng2024biscuitscaffoldingllmgeneratedcode} approach.    
\end{itemize}

\begin{figure}
    \centering
    \includegraphics[width=1\linewidth]{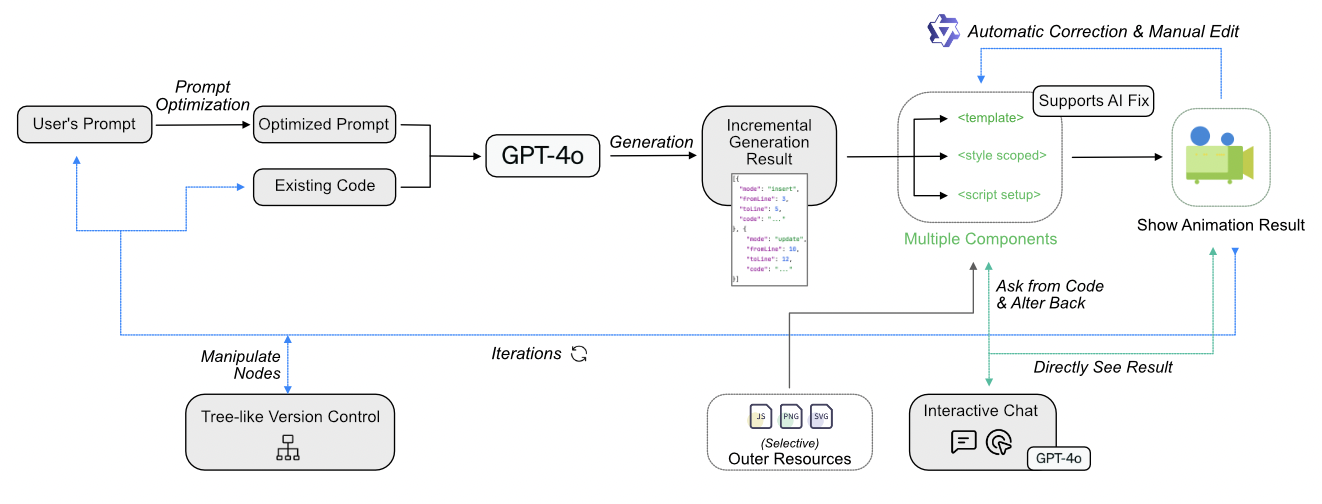}
    \caption{Workflow of AnyAni system}
    \label{fig:workflow}
\end{figure}

The AnyAni system also includes other panels, such as the Resources Panel, Settings Panel, Prompt Optimization Panel, etc. All panels are integrated together to form a complete workflow, as shown in Fig.~\ref{fig:workflow}.

\subsection{Structural and Incremental Generation}

Existing research has limited insights on how to improve the user experience of multiple rounds of iterations over complex generated code. Non-linear version control and real-time representation, as demonstrated by Spellburst~\cite{10.1145/3586183.3606719}, are important components and have also been implemented in our system. However, beyond this, there are more considerations about the quality, readability and usability of generated content. Over our own generation attempts and the formative study, we extracted these shortcomings:

\begin{itemize}
    \item \textbf{Overly simplistic and generic demos.} When complex creative programming generation tasks are performed using existing large models, the results generated are often unsatisfactory and oversimplified, as indicated by U2 and U4. Thus, iteration is a necessary task.
    \item \textbf{Lazy and omitted LLM Output.} After multiple rounds of iteration, the new output code will often omit parts that closely resemble those in the preceding version. This results in generated code that will not be directly usable, causing user experience issues. Common forms of lazy outputs include: ``\texttt{<!-- Simplistic representation -->},'' ``\texttt{/* add style here */},'' ``\texttt{/* code omitted */},'' ``\texttt{// other lines of code unchanged},'' etc. These identifiers, rather than the original code, are displayed in the answer. Existing solutions that use large models to assist in creative programming, such as Spellburst\cite{10.1145/3586183.3606719}, have not effectively addressed this issue.
    \item \textbf{Excessively long output.} After iterations, even though the LLMs are sometimes lazy, the generated answers are often still excessively long. Therefore, understanding the changes and seeking for valuable thinking can often become difficult (U1, U4).
    \item \textbf{Erroneous code.} Occasionally, the code generated by large models is entirely non-executable.
\end{itemize}

These shortcomings also align with DP~\ref{dp-1} and DP~\ref{dp-3}, and are attempted to be solved via the structural and incremental generation feature of the AnyAni system. Our paradigm may also provide valuable insights for the generation and iteration of code in many other creative programming tasks.

\subsubsection{Structural regulations}

We conducted extensive testing on the front-end code output (HTML, CSS, Vue, etc.) generated by LLMs, and found that the GPT-4o model possesses a fairly deep understanding of the parameters in these programming languages and can generate syntactically correct code. However, the visual outcomes it produces are not always optimal, and it doesn't consistently meet all user requirements. Additionally, issues such as laziness during generation and inconsistency in style are prominently observed. 

\begin{figure}[htbp]
    \centering
    \includegraphics[width=0.7\linewidth]{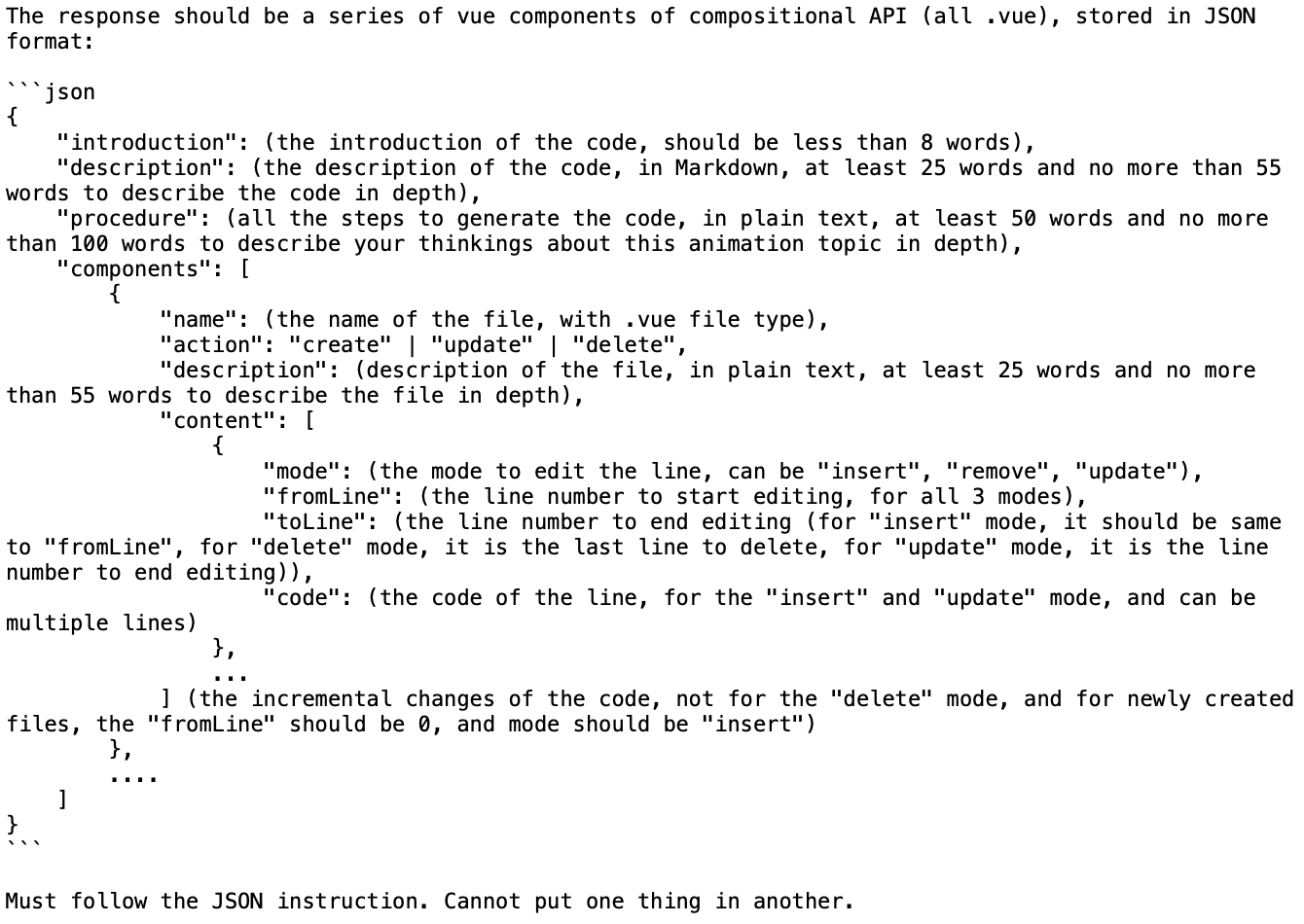}
    \caption{Structural constraints in the system prompt}
    \label{fig:system-prompt}
\end{figure}

Thus, we designed a system prompt paradigm requiring GPT to produce structured JSON-based outputs, enabling multiple iterative modifications. This pattern also reduces the code generation pressure even further by grouping the generated code into multiple discrete modules, which also reduces token consumption. In addition, our prompt incorporates Chain of Thought (CoT)~\cite{wei2023chainofthought} techniques to ensure the generated output more closely aligns with user requirements. The definition of the structure is shown in Fig.~\ref{fig:system-prompt}.

\subsubsection{Incremental generation} \label{incremental-gen}

\begin{figure}[htbp]
    \centering
    \includegraphics[width=0.66\linewidth]{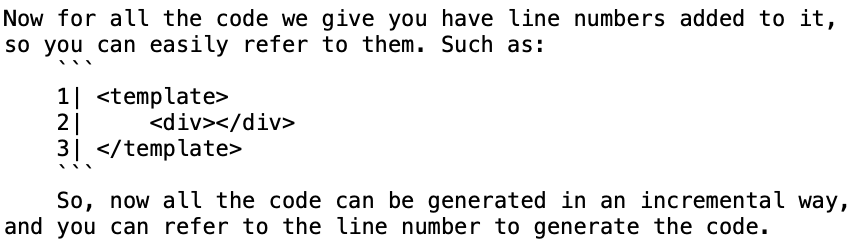}
    \caption{Incremental generation prompt}
    \label{fig:incremental}
\end{figure}

However, after the aforementioned optimizations, we still found that the model's tendency to produce lazy outputs could not be avoided through prompt engineering, especially when users are working from longer existing code. Thus, we introduced a straightforward but highly effective method of incremental generation, as shown in Fig.~\ref{fig:incremental}.

Before transmitting the code to the large model, we first use JS-Beautify~\footnote{\url{https://github.com/beautify-web/js-beautify}} to format the code and mark each line with the format ``(No.) |'' at the beginning. This allows for precise identification and ensures that only these lines are modified when changes are made. Each modification has three possible types: insert, remove, and update. Each modification type has \texttt{fromLine} and \texttt{toLine} properties (in the case of insert modifications, \texttt{fromLine} is equal to \texttt{toLine}), indicating that only these lines are to be altered. Finally, the effective changes are obtained, and based on the line number markers, the code in these lines is identified and modified. Here, each modification also queries the \texttt{(Line \#) |} marker first and then makes the change, ensuring that multiple modifications to adjacent lines do not result in errors. Finally, these line numbers are removed, yielding code components that can be run directly. Overall, this method is stable and effective, saves token space, further ensures consistent generation styles, and achieves good results.

However, in practical tests using GPT-4o, there is still a small chance of insertion errors occurring, especially for very long code (more than 200 lines). After numerous trials, we identified three common scenarios: \textbf{(1)} CSS and JavaScript content was not correctly added to the \texttt{style} or \texttt{script} tag. \textbf{(2)} A CSS selector was added within another CSS selector. \textbf{(3)} A newly added CSS rule (such as an \texttt{animate} rule) was not categorized under any CSS selector, and in most cases, this rule should have belonged to the next CSS selector. Once we have identified these issues, we can use PostHTML~\footnote{\url{https://posthtml.org}} and PostCSS~\footnote{\url{https://postcss.org}} to correct them, thereby reducing the likelihood of errors, along with other types of code errors that may be caused by generation.

\subsection{Code Understanding Assistance for Generated Code} 

\begin{figure}[htbp]
        \centering
	\begin{minipage}[t]{0.29\linewidth}
		\centering
		\includegraphics[width=0.99\linewidth]{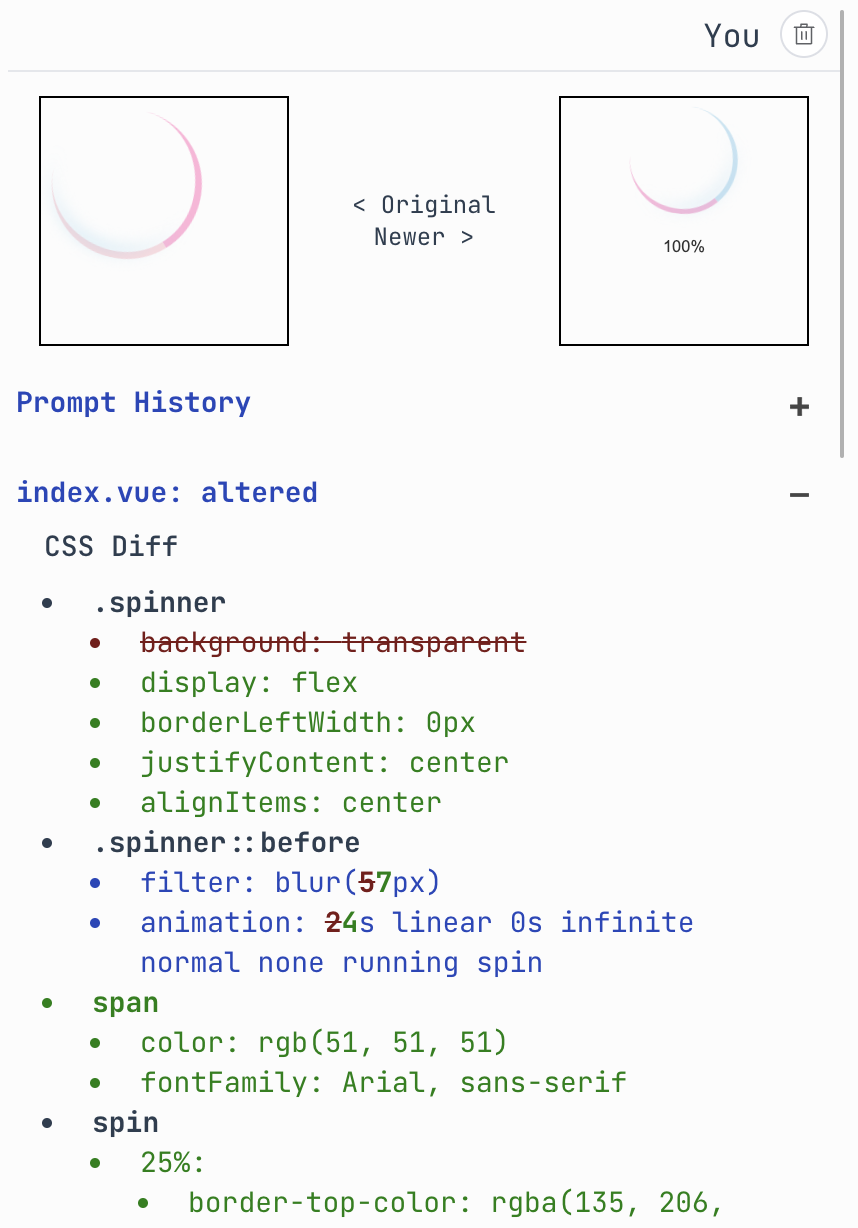}
		\caption{Comparing code difference}
		\label{fig:diff-comp}
	\end{minipage}
        \hspace{0.1cm}
 \begin{minipage}[t]{0.24\linewidth}
		\centering
		\includegraphics[width=0.99\linewidth]{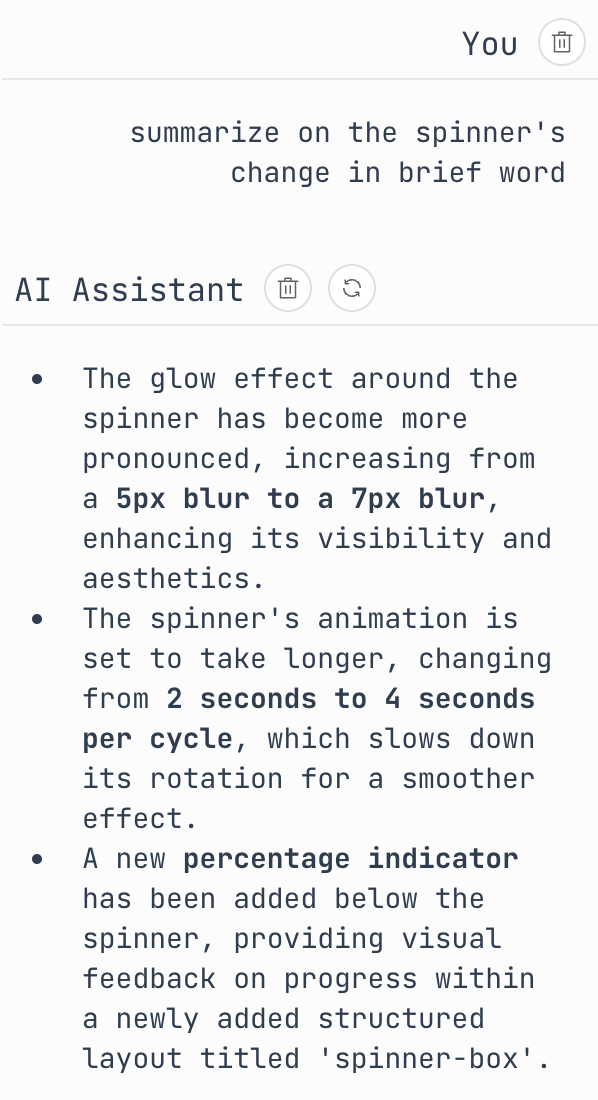}
		\caption{Summarizing difference}
		\label{fig:diff-sum}
	\end{minipage}
        \hspace{0.1cm}
	\begin{minipage}[t]{0.33\linewidth}
		\centering
		\includegraphics[width=0.99\linewidth]{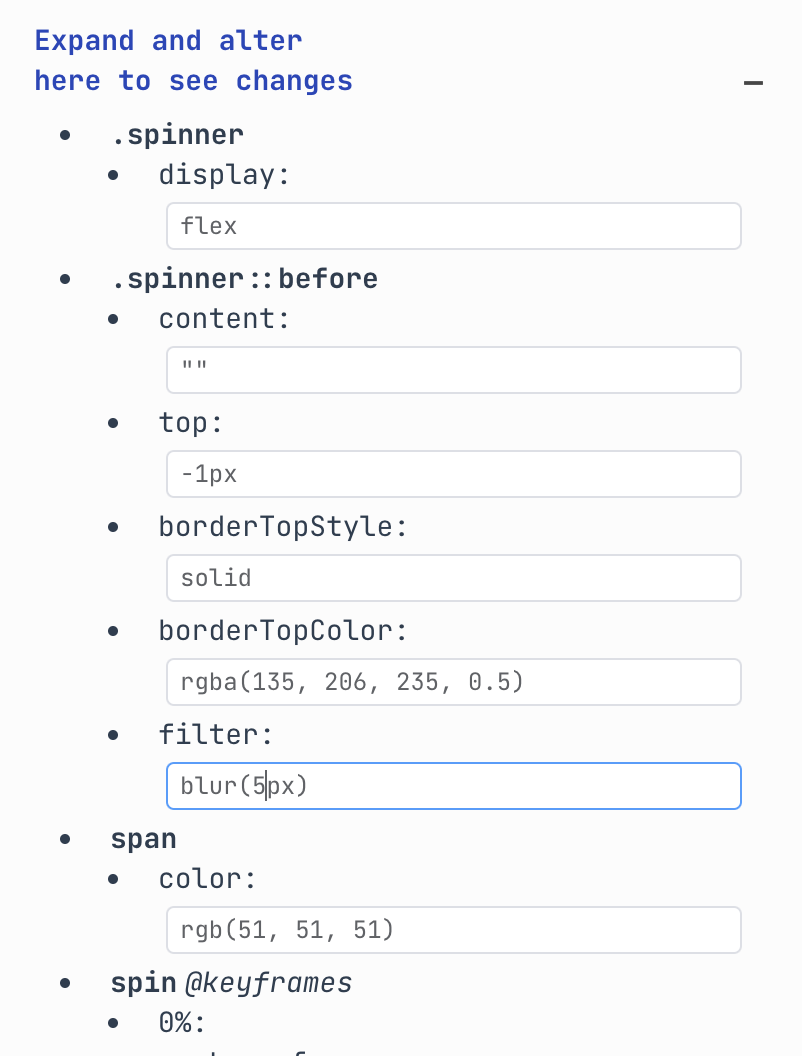}
		\caption{Learning-by-play support}
		\label{fig:learn-by-play}
	\end{minipage}
\end{figure}

After formative study, we noticed that adequate code understanding support, especcially the understanding of newly generated code will be an important factor (DP~\ref{dp-2}). Thus, difference comparation (diff) could be essential, as participants in the formative study (U2, U6) also pointed out that traditional diff formats could not meet their needs. Therefore, after exploring, we implemented a semantic code difference analysis interface~\ref{fig:diff-comp}, which is constructed through AST analysis. In addition, we integrate LLMs into diff analysis by transforming the results of diff analysis into computer-readable structured text as a background, alongside the user’s prompt history, enabling interactive diff analysis so that users can continue to ask LLM for summarizing or further learning and understanding~\ref{fig:diff-sum}. Furthermore, inspired by the concept of “Learn by Play~\cite{cheng2024biscuitscaffoldingllmgeneratedcode},” we used parsed AST object data along with the current dialogue content to retrieve relevant parameter lists in real time, in Fig.~\ref{fig:learn-by-play}. Users can interactively modify these parameters directly in the Interactive Chat Panel, and all changes are immediately reflected in the Show Panel upon submission. In this way, we further empower users to gain a faster understanding of the code, better meeting the requirements of DP~\ref{dp-2}.

\subsection{Automatic Video-based Correction}

Although existing research has made strides in the domain of automated repair, the focus has been predominantly on syntactic correctness. For instance, studies such as~\cite{10.1145/3641399.3644113} emphasize ensuring code executes without errors. Conversely, works like~\cite{liu2024logomotion} consider visual aspects but are limited to detecting the structural positioning of elements. Consequently, these approaches fall short in assessing more complex tasks, such as the creation and evaluation of animation elements from scratch, and with regard to verifying whether the animation effect itself meets specified requirements. In the meantime, recent researches in multi-modal models, such as Qwen2.5-VL~\cite{bai2025qwen25vl} and CogVLM2-Video~\cite{hong2024cogvlm2}, have emerged with capabilities to comprehend videos directly, including temporal and spatial information, which are very critical for animation effect understanding. These innovative models address the limitations of the previous generation LLMs that merely supported image understanding or relied on image-based video comprehension, exemplified by models like GPT-4V~\cite{2023GPT4VisionSC}. 

To analyze animations, we first record animation results based on the web technology stack into videos that can be directly used as input sources for a video understanding language model. We also input all prompts and descriptions of the generated animations from the current branch to examine whether the animations fully meet user requirements. If not, the specific unmet parts are indicated by the video understanding model, and are then sent as prompts to the code generation model for incremental generation. In complex tasks, to effectively achieve user expectations, this video understanding-based analysis process can be executed multiple times, progressively optimizing the generated results. Traditional automatic correction based on code quality remains essential; however, for tasks focused on visual and dynamic effects, analysis based directly on video understanding may be crucial. To summarize, this feature further resolves challenges behind DP~\ref{dp-1} and DP~\ref{dp-3}.

\subsection{Implementation}

The implementation of AnyAni was developed by the first author of this paper using Vue~\footnote{\url{https://vuejs.org}} and Typescript~\footnote{\url{https://www.typescriptlang.org}}. It utilized Element Plus~\footnote{\url{https://element-plus.org}} and Tailwind CSS~\footnote{\url{https://tailwindcss.com}} as UI support libraries, incorporated Vue Flow~\footnote{\url{https://vueflow.dev}} and Dagre~\footnote{\url{https://github.com/dagrejs/dagre}} for constructing tree structure modules, employed Monaco Editor~\footnote{\url{https://microsoft.github.io/monaco-editor}} for building code-related panels, leveraged Golden Layout~\footnote{\url{https://golden-layout.com}} for highly customizable page layouts, and integrated OpenAI's GPT-4o~\footnote{\url{https://openai.com/index/gpt-4o-system-card}} as the selected LLM model for generating code, explanations, and conversation content. For video-based analysis and automatic fixing, Dom-to-Image~\footnote{\url{https://github.com/tsayen/dom-to-image}} is used to transfer web animation to images, the Media-stream Recording API of JavaScript is utilized generate playable WebM video, which are then trans-coded into a MP4-formatted video using FFmpeg.js~\footnote{\url{https://github.com/Kagami/ffmpeg.js}} due to compatibility issues, and Qwen 2.5-VL~\footnote{\url{https://qwenlm.github.io/blog/qwen2.5-vl-32b}} is the base model for video analysis.

The system we developed encompasses every step from begin to end. This includes optimizing prompts, LLM code generation, manual code alteration, version controlling, questioning, and analysis. The animation effects are created in pure CSS, supported by multiple Vue components, and lively rendered on the page with iframes, using Vue's global build version.
\section{Evaluation}

During the evaluation process, we focused on determining whether front-end developers, who are not experienced animation creators, could create web animation effects more conveniently through the AnyAni system, and achieve better understanding and learning. We are also interested in how people use this system in practice.

\subsection{Method}

We recruited a group of programmers with extensive front-end development experience (N=9, 3 females and 6 males, aged 21 to 29) for user study, referred to hereafter as E1-E9. These individuals were recruited from our professional networks as well as a local university's social networks and mailing lists, and all recruited participants had at least two years of front-end development experience, and had worked on several real-world projects; although all participants had experience using large language models, their degree of integrating these models into front-end development and creative programming varied. About the users' proficiency in web-based animation creation, the majority (67\%) of them are not experienced animation creators, and the minority (33\%, E1, E2, E9) are also included in the process for acquiring expert opinions.

The interview sessions were organized by the first author of this paper and conducted in the form of one-on-one discussions. Each interview lasted approximately 60-75 minutes, and participants received compensation equivalent to about \$20 in local currency. 

The study involves collecting participants' demographic and technical background, demonstrating the AnyAni system, having them complete two animation coding tasks (using both AnyAni and a WebStorm+GPT-4 baseline in counterbalanced order), followed by questionnaire responses, log analysis, and semi-structured interviews to evaluate system performance.

The two tasks are:
\begin{enumerate}
    \item \textbf{Task 1.} Implement a loading animation that the user likes. This task can test the system's support for idea building, generation, and exploratory programming from scratch.
    \item \textbf{Task 2.} Implement an effect that animates an existing CSS-based submarine. By utilizing an existing long code base, this task can test our system's performance in complex scenarios while providing sufficient support for code comprehension.
\end{enumerate}

\subsection{Findings}

\subsubsection{System Usability} \label{usability}

After completing the assessment study, we asked participants to immediately complete the NASA Task Load Index (NASA-TLX) questionnaire \cite{nasa-tlx} for our system and the Post-Study System Usability Questionnaire (PSSUQ) \cite{pssuq} for both our system and the baseline (WebStorm + GPT-4o).

\begin{figure*}[htbp]
    \centering
    \includegraphics[width=0.95\linewidth]{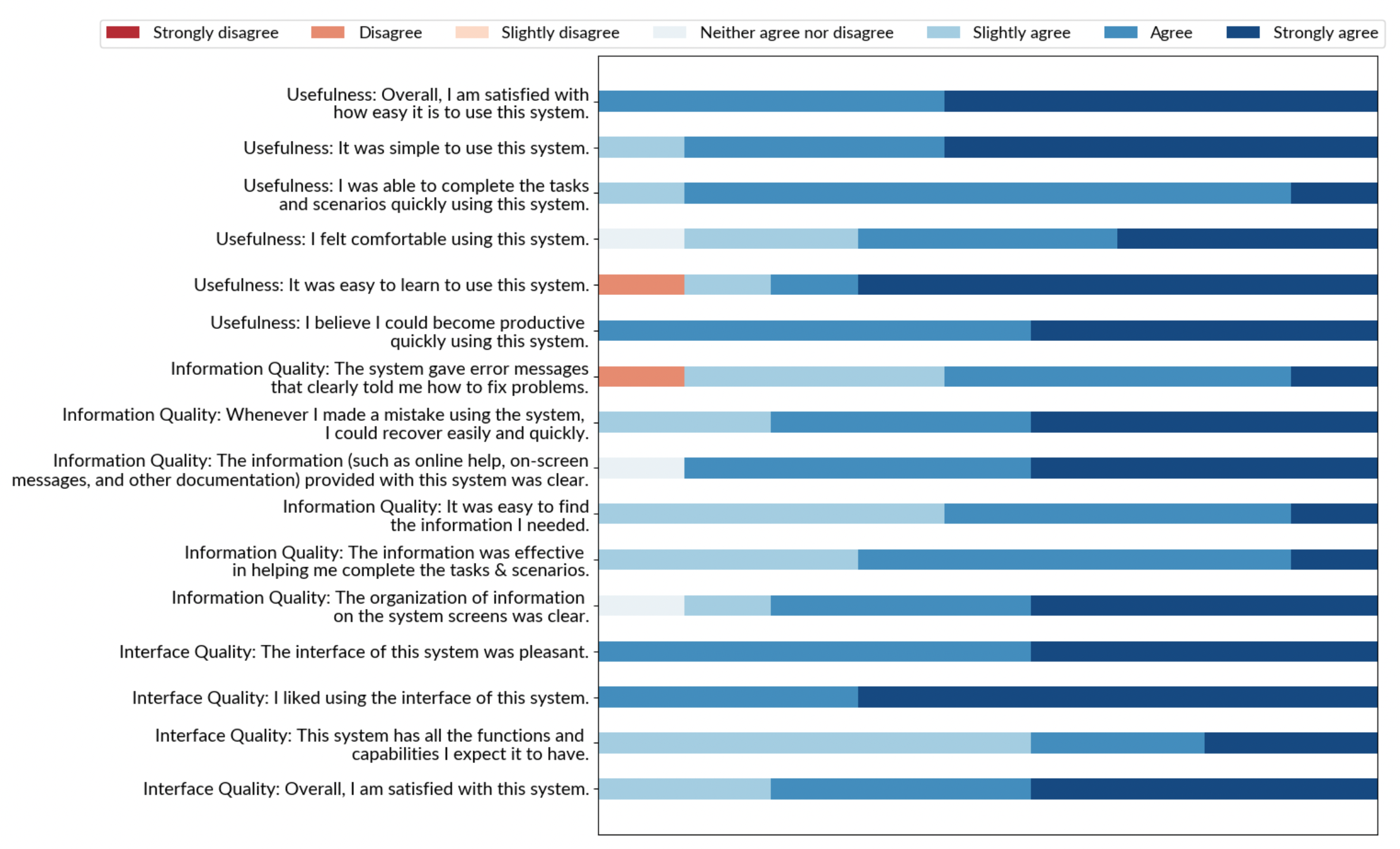}
    \caption{Participants' response to PSSUQ questionnaire}
    \label{fig:pssuq}
\end{figure*}

In the PSSUQ (7-point scale \cite{doi:10.1080/10447318.2002.9669130}) survey results displayed in Fig \ref{fig:pssuq}, we found that the overall usability score for the AnyAni system was 6.11/7, indicating that our system exhibited a high level of usability, and received recognition from our participants. A large proportion of ``Strongly Agree'' and ``Agree'' responses in the illustration further visually corroborates this.

\subsubsection{Comparison with the GPT-4o baseline}

To compare our system with GPT-4o baseline, we also asked participants to fill out two NASA-TLX scale based on their experience. This allows us to compare our results against the baseline to identify advantages and potential shortcomings. In this study, we followed NASA-TLX scale standards~\cite{nasa-tlx} to define the scoring for each performance item, where 0 indicates complete success and 20 indicates complete failure. 

\begin{figure}[htbp]
    \centering
    \includegraphics[width=0.99\linewidth]{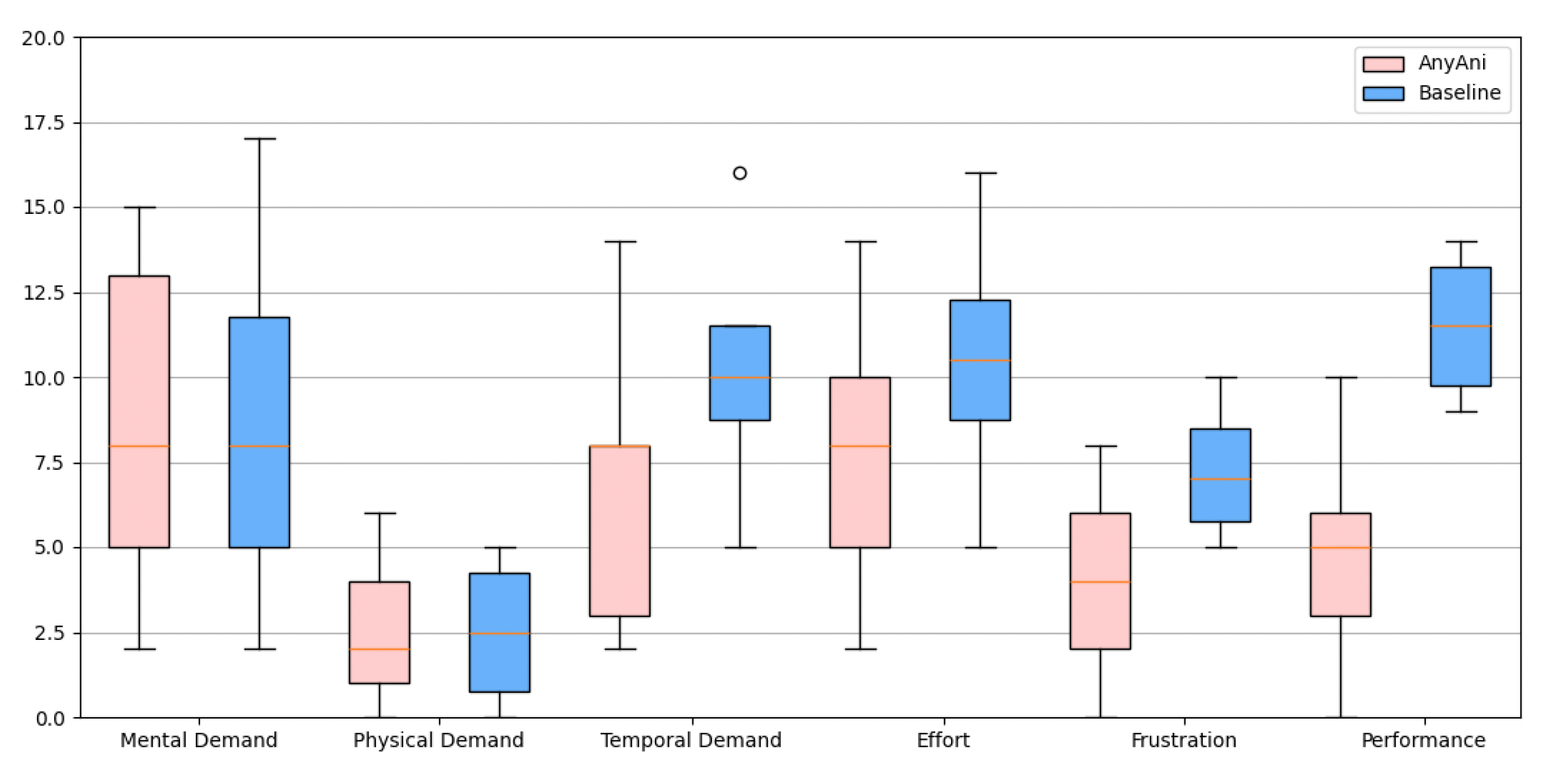}
    \caption{Comparison of NASA-TLX Cognitive Load Questionnaire results between AnyAni and GPT-4o baseline}
    \label{fig:nasa-tlx-comp}
\end{figure}

From Fig \ref{fig:nasa-tlx-comp}, it is evident that our system visibly outperforms the baseline across four indicators (with lower scores indicating better evaluations), although our system scores similarly to the baseline system in terms of mental and physical demand due to the nature of the animation effect creation task. 

\begin{itemize}
    \item In terms of \textbf{temporal demand}, our system has an average score of 7.22 (SD=4.31) compared to the baseline system's 10.25 (SD=3.89), indicating that users can achieve similar effects in a shorter time using our system.
    \item Regarding \textbf{effort}, our system's average score is 8.11 (SD=3.81) while the baseline system's average is 10.5 (SD=3.91), suggesting that users can attain the desired effects with less effort when using our system.
    \item In terms of \textbf{frustration}, our system has an average score of 4.0 (SD=2.58) whereas the baseline system scores 7.25 (SD=1.92), demonstrating that our system effectively reduces user stress in tasks that may cause anxiety, leading to more satisfactory outcomes (P-value=0.0622).
    \item As for \textbf{performance}, our system's average score is 4.44 (SD=2.71) compared to the baseline system's 11.5 (SD=2.06), indicating significantly better user creation results using our system (P-value=0.0013).
\end{itemize}

We also compared the scores from experts (E1, E2, E9) and non-experts, and found no significant difference, which means that our system had lower requirements for users' existing knowledge and may facilitate learning for non-expert users.

In our user study, many participants indicated that they often felt frustrated when using the native GPT interface for development and needed to invest a significant amount of effort to achieve the desired results.

User feedback highlights some common points of dissatisfaction with GPT. For instance, E1 remarked: \textit{``One thing I dislike about GPT is that it generates too much output, making it quite laborious to read line by line. This makes me feel very exhausted and lose interest in reading. Moreover, it often provides a lot of useless information, requiring me to filter it out myself. If it outputs a large block of code, I do not want to read it at all.''} E2, E4, and E6 provided similar feedback. In addition, E4 pointed out that when using the GPT interface, it is often difficult to clarify what specific content the system has modified. E4, E5, and E9 also expressed dissatisfaction with GPT's habit of truncating long code outputs, which prevents them from directly using the results, requiring them to first understand the edited parts and then manually insert them in the appropriate places to make sure that the code can run correctly. E4 revealed that this made her feel insecure, whereas using the AnyAni system eliminates this concern. E3 similarly mentioned that it is difficult to track and follow modifications when using GPT for multi-round iterations.

After experiencing the interface of AnyAni, many participants also found the interaction mode of GPT to be less suitable for this task compared to that of AnyAni. E1, E4, and E6 indicated that they were unable to directly assess the effectiveness of the generated code because the GPT system does not provide a preview of the results. In contrast, while using the AnyAni system, E4 mentioned, \textit{``Now I can intuitively see the implemented effect.''} Following code generation, participants are also pleasant about the subsequent interaction process within the AnyAni system. For instance, E6 remarked, \textit{``It (AnyAni) allows you to interactively modify parameters such as color and style.''}

Furthermore, participants commented on the possible user-friendliness of GPT for novice users. E9 pointed out that if he imagined himself as a novice user, \textit{``the biggest challenge in using GPT would be how to clearly describe the desired effects and design appropriate prompts to achieve the expected results.''} Moreover, E4 emphasized that due to the limitations of the linear dialogue interface, she found it challenging to rapidly obtain understandable explanations of specific code, which increased her cognitive load.

\subsubsection{Users' specific evaluations of the system} \label{eval}

Evaluation studies revealed high usability and efficacy of AnyAni: 100\% of participants agreed/strongly agreed on interface intuitiveness, while 78\% reported rapid learnability (Fig.~\ref{fig:pssuq}). Users praised the co-creative workflow, with E8 noting \textit{automated code fixes surpass pure AI outputs.''} Generated code quality received positive feedback, including E2's assessment of \textit{smooth effects''} and E9's appreciation for \textit{``non-disruptive incremental updates.''}

The advantages of the system summarized by users can be concluded in the following aspects:

\begin{itemize}
\item \textbf{Intuitive effect display \& interaction.} Users valued real-time code-effect synchronization, with E1 stating edits yield immediate feedback: \textit{I immediately know if this effect is useful.''} E7 emphasized direct code control (\textit{avoid arguing with the model''}), while E2 praised manual correction of AI imperfections. The triad of \textit{``code, model, and user collaboration''} (E7) enhanced workflow integration.

\item \textbf{Requirement refinement \& creativity support.} The system transformed abstract ideas into concrete outputs, surprising users with creative solutions (E1: \textit{``always gives surprises''}; E4: \textit{``inspires imagination''}). E5 noted effective interpretation of vague terms like \textit{``cool rotating effect,''} while E9 gained technical vocabulary aid as a non-native speaker.

\item \textbf{Multi-branch version tree navigation.} Multi-branch structures enabled iterative refinement, with E9 progressing \textit{``step by step''} and E4/E6 appreciating rapid historical state retrieval (\textit{``quickly locate/trace back''}).

\item \textbf{Precision code modification.} Incremental generation reduced manual code handling (E6: \textit{``no copying needed''}), outperforming raw GPT for long-code animations (E8). Pre-configured prompts lowered iteration effort (E7).

\item \textbf{Code comprehension \& learning.} Integrated explanations (E6: \textit{``AI Explain clarifies code''}) and diff comparisons supported principle learning. E5 likened it to \textit{``a practice-enabled tutorial,''} while E9 matched code to CSS selectors for intuitive understanding.
\end{itemize}

Participants have also noticed LLM error limitations but praised automated correction via the \texttt{AI Fix} feature. E7 found it \textit{convenient,''} while E6 deemed it \textit{beginner-friendly''}—forming a robust \textit{``write-fix-rewrite''} loop (E1). Error resolution mechanisms are detailed in Section~\ref{incremental-gen}.

\subsubsection{Analysis of user activity logs}

Additionally, while users engage with our system for research purposes, we automatically collect their click data and other usage statistics. This enables us to better evaluate user interactions with our system, providing valuable insights for the future development and assessment of AI-CSTs \cite{10152832}. All data are fully anonymous and free from any personal information, and participants are informed of this recording in advance.

\begin{figure}[htbp]
    \centering
    \includegraphics[width=0.99\linewidth]{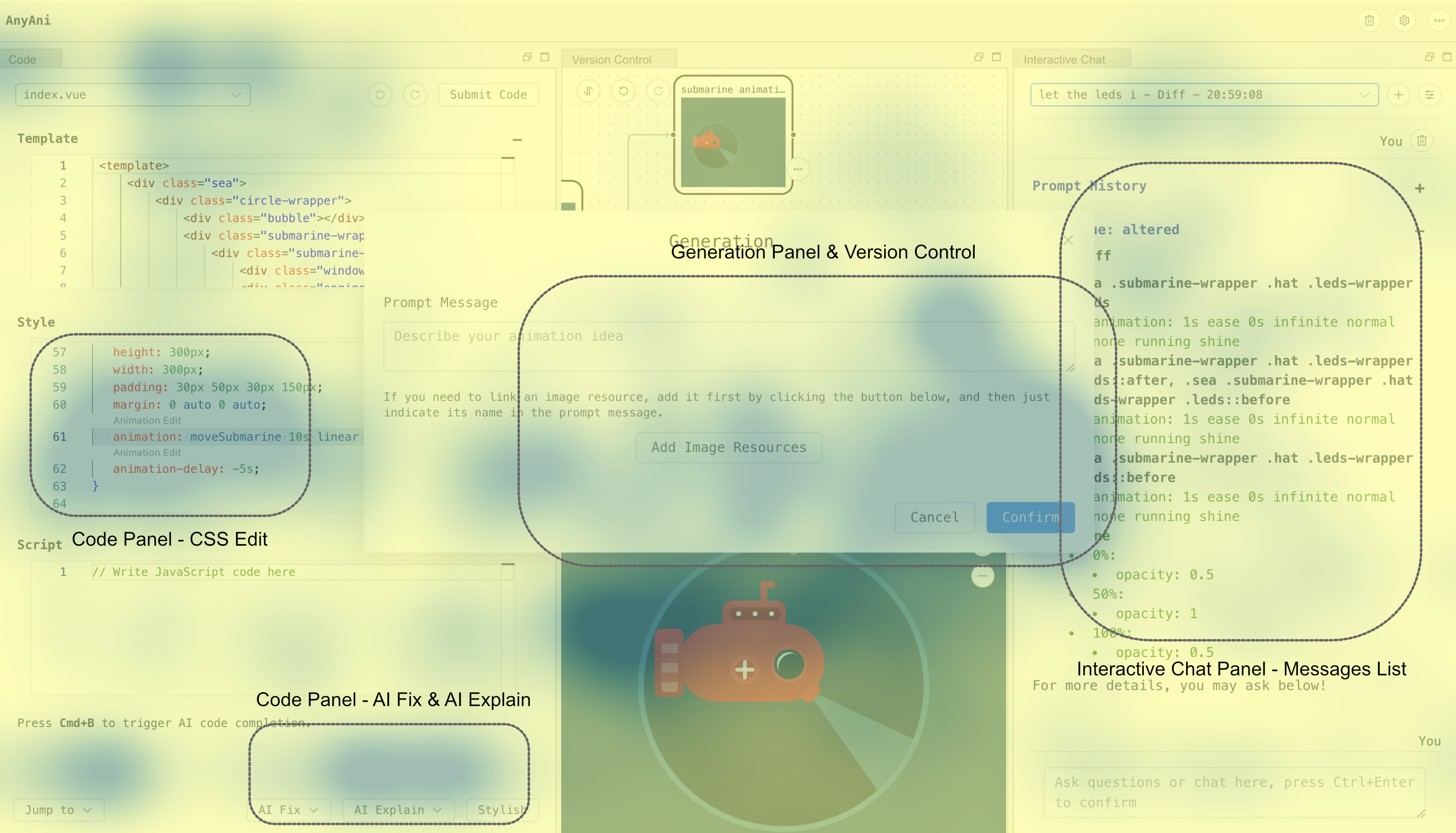}
    \caption{Usage record heat map (blue indicating more clicks and yellow indicating fewer clicks)}
    \label{fig:heatmap}
\end{figure}

We recorded users' click data on the interface and extracted the primary click hot-spots. We visualized the click frequencies by plotting a heat map using Matplotlib \cite{Hunter:2007}. By observing the heat map, we could clearly identify the areas of the interface with high-frequency interactions during user sessions, which provides robust support for further optimizing the user experience.

\begin{enumerate}
    \item The CSS Style Edit section of the Code Panel is frequently used as editing CSS code is undoubtedly the foundation for creating CSS animations.

    \item The AI Fix and AI Explain sections in the Code Panel are also frequently used. Since both buttons trigger pop-up menus, the high-frequency area are located above the buttons. This suggests that the AI Fix and AI Explain functionalities indeed enhance users' understanding and learning of code, so they are willing to utilize them frequently.

    \item The Generation Panel and the Version Control Panel (partially obscured) are also high-frequency areas, indicating that users extensively utilize our tool's generation functionalities and version management panel for generative and exploratory programming.

    \item Moreover, The Messages List in the Interactive Chat Panel is also frequently clicked, evidencing that users engage in substantial interactive operations while conversing with the model.
\end{enumerate}
\section{Discussion and Future Work}

Although our evaluation study with front-end developers has provided valuable insights into the effectiveness of AnyAni, there are still some limitations to consider. These limitations also provide a good opportunity for further future work. Additionally, we also recognize the potential ethical risks that may arise when using LLMs to assist in digital art creation.

\subsection{Medium of Animations}

The decision to prioritize CSS animations over JavaScript-based implementations reflects a deliberate design choice informed by our formative study. While CSS offers lower expressiveness than programmatic animation libraries, its declarative nature provides a constrained design space that proved more conducive to reliable AI generation and user comprehension, which is particularly critical for novice developers. However, as discussed in Section~\ref{anim-tools}, there are also many JavaScript-based animation effect libraries that can implement more complex animation effects more conveniently. Future research should attempt to fine-tune the model to enhance the system's support for JavaScript content and offer assistance for various JavaScript animation effect libraries.

Additionally, E1 indicated that when creating complex web animations such as skeletal animations in his actual job, he often ends up returning to Adobe After Effects~\footnote{\url{https://www.adobe.com/products/aftereffects.html}} and export in Lottie~\cite{lottie} format, which is a field where AI cannot participate in creation now. Therefore, we now have web animations that can be generated and freely edited, more complex but harder-to-edit generated animations based on video generation technology, and beautifully crafted, freely editable animations with lower AI integration made by traditional software. Methods to integrate these three paths to support the AI-aided creation of more complex animation effects while retaining their editability undoubtedly remain a significant focus for future research.

\subsection{Research Scope}

Our scope of research remains relatively limited at this stage, as the existing system was predominantly evaluated by front-end developers, finding that both front-end developers with expertise in web animation creation and non-experts were generally satisfied with the system. However, the group interested in or attempting to generate web animation effects could be much broader. For instance, artists, UX designers, project managers and full-stack engineers, who are non-specialist front-end developers, might also be interested in this field. Furthermore, we can anticipate that some of these individuals might be eager to learn the necessary skills from scratch. Therefore, future research should expand the sample range to encompass individuals with more diverse backgrounds in order to better assess the system's performance and usability, particularly for those without any knowledge in front-end development, and explore whether the AnyAni system can help them achieve understanding and learning objectives.

Meanwhile, we also note that large-scale creative work usually requires collaborative efforts~\cite{10.1145/1323688.1323689,10.1145/3586183.3606719}. Therefore, we can further explore the usage of AnyAni's tree management and WYSIWYG editing paradigms in collaborative creation stages, and actively investigate more possibilities of multi-user collaboration tools in creative programming in future work.

\subsection{Future of Co-creation with AI}

The AnyAni system contributes to emerging computer-aided design (CAD) frameworks for human-AI collaboration by demonstrating how generative models can participate as process-oriented partners rather than outcome generators. Our implementation of incremental code generation with contextual awareness extends traditional ``prompt-and-replace'' interactions~\cite{liu2022design} into a collaborative dialogue where human and AI contributions become interleaved artifacts in a shared design history. This approach resonates with Kantosalo and Toivonen's~\cite{Kantosalo2016ModesFC} concept of ``task-divided co-creativity,'' where AI handles mechanical implementation while humans focus on high-order creative decisions.

Three key insights emerge for future co-creation systems:

\begin{itemize}
    \item \textit{Temporal Context Preservation}: The version tree's branching structure addresses a critical limitation identified in our formative study, that is LLMs' inability to maintain consistent context across iterations. By enabling selective context inheritance through parent node relationships, we provide a mechanism for managing ``creative micro-versions''~\cite{nicholas2022creative} in exploratory programming.
    \item \textit{Bidirectional Accountability}: Unlike typical chatbot interfaces, AnyAni's diff analysis and parameter manipulation features create mutual observability between human and AI contributions. This transparency is crucial for maintaining user agency – a recurring concern in AI-assisted creativity tools~\cite{10152832}.
    \item \textit{Skill-Adaptive Abstraction}: The system's layered interface (from visual parameter controls to raw code editing) suggests a promising direction for adaptive UIs that respond to users' growing expertise. Future systems could employ machine learning to detect skill progression and automatically adjust abstraction levels, which is a concept suggested in code education research~\cite{kazemitabaar2024codeaid}.
\end{itemize}

Looking ahead, three research directions demand attention: \textit{(1)} Developing multi-modal interfaces that combine text prompts with visual timeline manipulation for complex animations; \textit{(2)} Investigating collaborative workflows where multiple humans and AI agents co-author animation sequences; and \textit{(3)} Establishing evaluation metrics for AI-assisted creativity that balance technical execution with artistic merit. By addressing these challenges, the CAD community can advance generative AI from a productivity tool to a true creative partner in interface development.

\section{Conclusion}

This paper presents AnyAni, a novel system that allows users to create, explore, and understand web animation effects using a human-in-the-loop approach leveraging LLM. Evaluation studies demonstrate the system's usability and effectiveness in reducing development time, enhancing code comprehension, and fostering exploratory programming. Users appreciate the system's intuitive interface, real-time preview, and interactive code manipulation capabilities. We are excited to introduce the AnyAni system's comprehensive workflow, the concepts, and paradigms behind it, which may help further explore the possibilities of enhancing human creativity.

\bibliographystyle{ACM-Reference-Format}
\bibliography{main.bib}

\appendix
\appendix

\section{Participants In Evaluation Study} \label{participants}

In the section of \textit{programming experience}, we present participants' duration of programming practice, duration of front-end development, current level in creating web animations, and frequency of using large models to assist in development. We represent novice users as NOV, intermediate level or frequency as MID, and expert users as EXP. 

\begin{table}[htbp]
\resizebox{0.99\linewidth}{!}{
\begin{tabular}{ccccc}
\hline
   & Gender & Age & Career Experience           & Experience \\ \hline
E1 & Male   & 29  & IT Company      & 8 / 4 / EXP / NOV      \\
E2 & Male   & 23  & Graduate Student    & 4 / 3 / EXP / EXP      \\
E3 & Male   & 25  & FinTech Company & 6 / 3 / NOV / EXP      \\
E4 & Female & 25  & HCI Researcher              & 3 / 3 / MID / NOV      \\
E5 & Female & 21  & Undergraduate Student       & 3 / 2 / MID / MID      \\
E6 & Male   & 22  & Graduate Student            & 4 / 2 / NOV / EXP      \\
E7 & Male   & 20  & Undergraduate Student       & 3 / 2 / NOV / EXP      \\
E8 & Female & 23  & Graduate Student            & 5 / 3.5 / MID / EXP    \\
E9 & Male   & 26  & FinTech Company & 9 / 4 / EXP / EXP      \\ \hline
\end{tabular}
}
\caption{The list of participants in our user study.}
\label{tab:my-table}
\end{table}

For example, E1 has 8 years of programming experience, 4 years of front-end development experience, is an expert in web animation creation, and although he has experienced LLMs, he has not integrated them into his programming workflow yet, so he is described as \texttt{8 / 4 / EXP / NOV}.

\end{document}